\documentclass[conference,a4paper]{IEEEtran}
\IEEEoverridecommandlockouts
\usepackage{cite}
\usepackage{amsmath,amssymb,amsfonts}
\usepackage{algorithmic}
\usepackage{graphicx}
\usepackage{textcomp}
\usepackage{xcolor}
\usepackage[caption=false]{subfig}
\usepackage{circuitikz}
\usepackage[hidelinks]{hyperref}
\def\BibTeX{{\rm B\kern-.05em{\sc i\kern-.025em b}\kern-.08em
    T\kern-.1667em\lower.7ex\hbox{E}\kern-.125emX}}

\begin{document}

\title{VSC-HVDC setpoint adjustment for maximum \\grid utilisation under voltage constraints
\thanks{This work has been carried out within the framework of the SPS4SE project, which is funded by the Swedish Energy Agency.}
}

\author{\IEEEauthorblockN{1\textsuperscript{st} Gabriel Malmer}
\IEEEauthorblockA{\textit{Industrial Electrical Eng. \& Automation} \\
\textit{Lund University}\\
Lund, Sweden \\
gabriel.malmer@iea.lth.se}
\and
\IEEEauthorblockN{2\textsuperscript{nd} Emil Hillberg}
\IEEEauthorblockA{\textit{Electric Power Systems} \\
\textit{RISE Research Institutes of Sweden}\\
Gothenburg, Sweden \\
emil.hillberg@ri.se}
\and
\IEEEauthorblockN{3\textsuperscript{rd} Olof Samuelsson}
\IEEEauthorblockA{\textit{Industrial Electrical Eng. \& Automation} \\
\textit{Lund University}\\
Lund, Sweden \\
olof.samuelsson@iea.lth.se}
}

\maketitle

\begin{abstract}
Voltage-source converter high-voltage direct current (VSC-HVDC) links offer controllable active and reactive power output, making them a promising asset for emergency voltage support. This paper presents an analytical method for adjusting the active power setpoint of a VSC-HVDC station to maximise loadability during voltage-stressed conditions. By exploiting the geometric structure of converter capability limits, a closed-form expression for the optimal setpoint is derived under combined current and voltage constraints. The method requires local voltage measurements and an estimate of a wide-area voltage angle difference, making it suitable for real-time emergency control without global optimisation. Validation on the Nordic Test System confirms that the analytically predicted optimum is consistent with the setpoint yielding the highest loadability, and that adjustments of active power setpoint can yield a more-than-proportional increase in loadability. The results further indicate robustness to angle estimate uncertainty.
\end{abstract}

\begin{IEEEkeywords}
converter limits, corrective actions, emergency power control, SIPS, VSC-HVDC, voltage stability
\end{IEEEkeywords}

\section{Introduction}
The demand for transmission capacity is growing, making efficient grid use while maintaining secure system operation imperative. Fast corrective control actions, when properly designed, are among the tools that can achieve both targets. 

Meanwhile, the share of converter-interfaced resources is increasing, with voltage-source converter high-voltage direct current (VSC-HVDC) transmission as a notable example. Several methods have been proposed to exploit the IGBT-based transmission technology's attractive control capabilities to alleviate grid constraints. A recent review of HVDC-based congestion management strategies is presented in \cite{zhou2026review}. The strategies range from dynamic corrective control to corrective security-constrained optimal power flow (CSCOPF). However, most approaches focus primarily on thermal overload mitigation, while voltage constraints are often neglected. Some methods consider voltage constraints but stay well within HVDC capability limits, thus no trade-off arises between active and reactive power support \cite{sass2020automated}. In \cite{mandoulidis2022overview}, it is acknowledged that converter current limits must be respected in voltage instability emergency controls, and that active power reduces the reactive control margin, but no guidance is given on how to balance the two. Making the most of existing controllable resources is critical, particularly as voltage collapse tends to occur when those resources are either depleted or not fully utilised. 

Johansson et al. show in \cite{johansson1997avoiding} and \cite{johansson1999mitigation} that rescheduling a small, local generator's active and reactive power $(P_g,Q_g)$ can increase the system capability in a local area during voltage (load) instability. In particular, when the generator is armature current-limited, the operating point enabling the highest load can be expressed in terms of the angle difference $\delta$ between the local and a voltage stable, remote bus: $Q_g/P_g = \tan{\phi} = 1/\tan{\delta}$. This relationship is used as a control law in \cite{johansson1997avoiding}, and its stability benefits are demonstrated for a small system. 

\subsection{Contributions}

While previous work has related active and reactive power rescheduling to voltage-stability margins, a general setpoint rule for VSC-HVDC stations under current and voltage limits combined has, to the authors' knowledge, not been established. The main contributions are as follows: first, we show that the control law in \cite{johansson1997avoiding} is translation invariant and generalises to any circular capability limit in the $P$--$Q$ plane, including generator over-excitation limiters (OELs) and VSC-HVDC current and voltage limits, while accounting for phase reactor, filter, and transformer. Second, we derive a closed-form expression for the optimal active power setpoint $P_g^*$ under combined current and voltage constraints. Third, we discuss how the derived expression can be used within an emergency power control (EPC) scheme to mitigate voltage instability. Unlike OPF-based redispatch methods, which typically require a system-wide model and online optimisation, the proposed approach provides a setpoint rule based only on local voltage magnitude and an estimate of the wide-area angle difference $\delta$, enabling low-complexity implementation in a fast EPC scheme. Validation on the Nordic Test System confirms that the derived expression accurately identifies the setpoint yielding the highest loadability, and that the loadability increase can be significant.

\section{System Modelling and Problem Formulation}

Using the relations by Beerten et al. \cite{beerten2012generalized}, the current and voltage limits of a VSC-HVDC converter station (Fig.~\ref{fig:vsc_equivalent}) at the system bus $\underline{U}_s$ can be represented as circles in the $P$--$Q$ plane with centre $\underline{S}_0 = P_0 + jQ_0$ and radius $r$.

\subsubsection*{VSC-HVDC current limit}

\begin{equation}
\label{eq:curr_lim}
\begin{aligned}
\underline{S}_0^{(i)} 
    &= -U_s^2 
       \left(\frac{1}{\underline{Z}_f^* + \underline{Z}_{tf}^*}\right), \\
r_i &= U_s I_{c_{\max}}
       \left|\frac{\underline{Z}_f^*}
       {\underline{Z}_f^* + \underline{Z}_{tf}^*}\right|.
\end{aligned}
\end{equation}

\subsubsection*{VSC-HVDC voltage limit}

\begin{equation}
\label{eq:volt_lim}
\begin{aligned}
\underline{S}_0^{(v)} 
    &= -U_s^2 \left(\underline{Y}_1^* + \underline{Y}_2^*\right), \\
r_v &= U_s U_{c_{\max}} Y_2 .
\end{aligned}
\end{equation}

Here, $U_s$ is the voltage magnitude at the system bus, and the feasible region is the intersection of the two circle interiors. The admittances $\underline{Y}_1$ and $\underline{Y}_2$ are obtained from the impedances $\underline{Z}_{c}$, $\underline{Z}_{f}$ and $\underline{Z}_{tf}$ (see Fig.~\ref{fig:vsc_equivalent}) through a $Y$--$\Delta$ transformation \cite{beerten2012generalized}. The limits $I_{c_{\max}}$ and $U_{c_{\max}}$ denote the maximum converter current and voltage magnitudes, respectively, where $U_{c_{\max}}$ represents the pulse-width modulation (PWM) saturation limit. 

\section{Optimal Active Power Setpoint}

Johansson et al. \cite{johansson1997avoiding} show that the maximum load $\underline{S}_d = P_d$ in a radial system with a small local generator is obtained for
\begin{equation}
\label{eq:reduced_cap_curve}
\frac{Q_g}{P_g} = \tan \phi = \frac{1}{\tan \delta},
\end{equation}

where $\underline{S}_g = P_g+jQ_g$ is the power injection at $\underline{U}_s$, $\phi$ is the power factor angle of the injection, and $\delta$ is the voltage angle difference between $\underline{E}$ and $\underline{U}_s$ (see Fig.~\ref{fig:vsc_equivalent}). We show (see appendix) that an analogous optimality condition holds for any circular limit, with arbitrary centre $(P_0,Q_0)$:
\begin{equation}
\label{eq:opt_any_centre}
\frac{Q_g - Q_0}{P_g - P_0}
= \frac{1}{\tan \delta}.
\end{equation}

Given the centre point and radius of each circle, the intersection points $(P_x,Q_x)$, if they exist, can be computed analytically \cite{bourke1992circles}. Selecting the intersection point with the larger active power, the transition angles $\delta_i$ and $\delta_v$ can be defined as
\begin{equation}
\begin{aligned}
\delta_i &= 
\arctan\!\left(
\frac{P_x - P_0^{(i)}}{Q_x - Q_0^{(i)}}
\right), \\
\delta_v &= 
\arctan\!\left(
\frac{P_x - P_0^{(v)}}{Q_x - Q_0^{(v)}}
\right).
\label{eq:deltai_deltav}
\end{aligned}
\end{equation}

Combining the current and voltage constraints, the optimal active power setpoint $P_g^*$ can be expressed as the following piecewise function of $\delta$:
\begin{equation}
\label{eq:Pg_set}
P_{g}^*(\delta) = 
\begin{cases}
r_v \sin \delta + P_0^{(v)}, 
& \delta < \delta_v, \\[4pt]
r_i \sin \delta_i + P_0^{(i)} = r_v \sin \delta_v + P_0^{(v)}, 
& \delta_v \le \delta < \delta_i, \\[4pt]
r_i \sin \delta + P_0^{(i)}, 
& \delta \ge \delta_i.
\end{cases}
\end{equation}

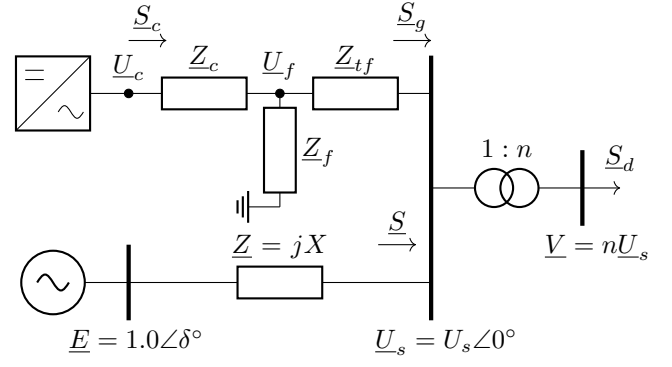
\begin{figure}[htbp!]
\centering
\begin{circuitikz}
\draw (1,0)
node[circ,label=above:$\underline{U}_c$] (Uc) {}
to[short] (0.5,0)
to[sacdc] (-0.5,0) (Uc) {}
to[generic,l=$\underline{Z}_{c}$, label distance=1pt](3,0)
node[circ,label=above:$\underline{U}_f$] (Uf) {}
to[generic,l=$\underline{Z}_{tf}$, label distance=1pt] (5,0)
node[] (Us_up) {};

\draw
(Uf) to[generic,l=$\underline{Z}_f$] (3,-1.5) 
(3,-1.5) node[ground,rotate=-90]{};

\draw[ultra thick] (5,-3) -- (5,0.5);
\node[below] at (5.2,-3) {$\underline{U}_s = U_s\angle0^\circ$};

\draw[ultra thick] (7,-1.75) -- (7,-0.75);
\node[below] at (7.2,-1.75){$\underline{V}=n\underline{U}_s$};

\draw (5,-1.25) 
to[oosourcetrans,l=$1:n$,label distance=-5pt] (7,-1.25);

\draw[->] (7,-1.25) -- (7.5,-1.25);
\node[above] at (7.5,-1.15) {$\underline{S}_d$};

\draw (0,-2.5) node[vsourcesinshape, rotate=90] (E) {} (E.south)
to[short] (1,-2.5)
to[generic,l=$\underline{Z} \ {=} \ jX$,label distance=1pt] (5,-2.5);

\draw[ultra thick] (1,-2) -- (1,-3);
\node[below] at (1.1,-3) {$\underline{E}=1.0 \angle \delta^\circ$};

\draw[->] (1,0.7) -- (1.5,0.7) node[midway,above] {$\underline{S}_c$};
\draw[->] (4.5,0.7) -- (5,0.7) node[midway,above] {$\underline{S}_{g}$};
\draw[->] (4.3,-2) -- (4.8,-2) node[midway,above] {$\underline{S}$};

\end{circuitikz}
\caption{Single line diagram of a local VSC-HVDC converter station connected to a strong grid (bus $\underline{E}$) through a tie reactance $X$, including converter impedance $\underline{Z}_c$, filter impedance $\underline{Z}_f$ and transformer impedance $\underline{Z}_{tf}$. The complex load $\underline{S}_d$ is connected to the system bus $\underline{U}_s$ through an idealised transformer with continuous tap control $1:n$.}
\label{fig:vsc_equivalent}
\end{figure}

For angles $\delta > \delta_i$, the optimum lies on the current limit. Similarly, for $\delta < \delta_v$, the optimum is constrained by the voltage limit. For $\delta_v \le \delta \le \delta_i$, the optimum is located at the intersection of the two limits, which is independent of $\delta$.

With $d$-axis (active power) priority, the reactive power $Q_g$, regulated via conventional AC voltage or droop control, is constrained by the most restrictive limit:
\begin{equation}
\label{eq:Qg_max}
\begin{aligned}
Q_{g}^{\max} = \min\bigg(& Q_0^{(i)}+\sqrt{r_i^2-(P_g-P_0^{(i)})^2}, \\
                & Q_0^{(v)}+\sqrt{r_v^2-(P_g-P_0^{(v)})^2} \bigg).
\end{aligned}
\end{equation}

\begin{figure}[htbp!]
    \centering
    \includegraphics[height=0.52\linewidth]{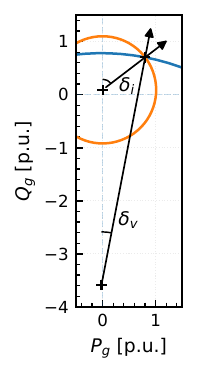}
    \hfill
    \includegraphics[height=0.52
    \linewidth]{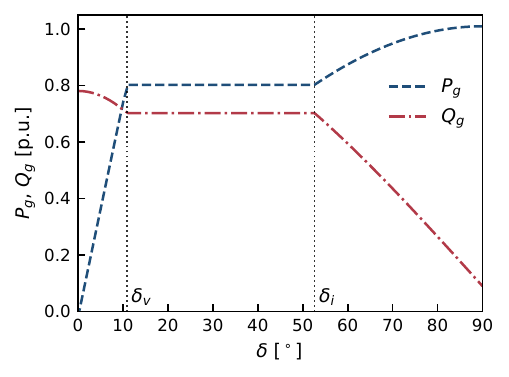}
    \caption{Left: $PQ$-capability chart at the system bus ($\underline{U}_s$) with current (orange) and voltage (blue) limits, and transition angles $\delta_i$ and $\delta_v$. $I_{c_{\max}} = 1.0~\mathrm{p.u.}$, $U_{c_{\max}} = 1.2~\mathrm{p.u.}$, $U_s = 1.0~\mathrm{p.u.}$ Right: Optimal active and reactive power ($P_g$, $Q_g$) as functions of $\delta$, showing transitions at $\delta_i$ and $\delta_v$.}
    \label{fig:ss_and_optimalPQ}
\end{figure}

Using $I_{c_{\max}} = 1.0~\mathrm{p.u.}$, $U_{c_{\max}} = 1.2~\mathrm{p.u.}$, and $U_s = 1.0~\mathrm{p.u.}$, in combination with the VSC converter parameters \cite{beerten2012generalized} in Table~\ref{tab:conv_data}, the resulting $PQ$-capability chart and optimal operating points $(P_g,Q_g)$ are shown in Fig.~\ref{fig:ss_and_optimalPQ}.

\begin{table}[htbp!]
\caption{VSC Converter Data}
\label{tab:conv_data}
\centering
\begin{tabular}{l c}
\hline
\textbf{Parameter} & \textbf{Value (p.u.)} \\
\hline
$\underline{Z}_c = R_c + jX_c$ 
    & $0.0001 + j0.1643$ \\

$\underline{Z}_f = 1/jB_f$
    & $-j11.27$ \\
    
$\underline{Z}_{tf} = R_{tf} + jX_{tf}$ 
    & $0.0015 + j0.1121$ \\

\hline
\end{tabular}
\end{table}

For modular multilevel converter (MMC) implementations, where filters are typically omitted ($\underline{Z}_f \rightarrow \infty$) due to the near-sinusoidal waveform output, the centre of the current limit coincides with the origin. This yields $Q_g/P_g = \tan{\phi} = 1/\tan{\delta}$, the same as in (\ref{eq:reduced_cap_curve}). Furthermore, if the resistive components of the VSC parameters are neglected (typically small, see Table~\ref{tab:conv_data}), $P_0^{(i)} = P_0^{(v)} = 0$ and (\ref{eq:Pg_set}) simplifies accordingly. The full expression is retained for generality.

\section{Implementation within an EPC framework}
\label{sec:epc_scheme}

The derived setpoint rule in (\ref{eq:Pg_set}) can be used as the basis for a real-time EPC system integrity protection scheme (SIPS), triggered under voltage-stressed conditions. Voltage-stressed conditions here refer to heavily loaded or post-contingency operating points lacking a long-term voltage stable equilibrium. The only time-varying quantities are $U_s$ and $\delta$: the former can be obtained from local measurements, and the latter via PMU-based wide-area measurements (WAMS) or the state estimator. For each control cycle, the following steps are executed:

\begin{enumerate}
\item \textit{Measurement and estimation:}  
The system bus voltage magnitude $U_s$ is measured locally. The voltage angle difference $\delta$ between the local bus and a remote, strong area is obtained from the WAMS or state estimator.

\item \textit{Parameter update:}  
Based on the measured $U_s$, the centres and radii of the current and voltage limits are updated according to (\ref{eq:curr_lim})--(\ref{eq:volt_lim}), and the transition angles $\delta_i$ and $\delta_v$ are calculated from (\ref{eq:deltai_deltav}).

\item \textit{Optimal setpoint computation:}  
Based on the measured $\delta$, the operating regime (voltage-limited, intersection, or current-limited) is identified, and the optimal active power setpoint $P_g^*$ is computed from (\ref{eq:Pg_set}).

\item \textit{Setpoint adjustment:}  
When the SIPS is triggered upon voltage instability detection, $P_g^*$ is sent as an additional input signal to the HVDC outer-loop control for the $d$-axis current reference $i_d^\mathrm{ref}$. The logic between $P_g^*$ and $i_d^\mathrm{ref}$ (and thus $P_g^\mathrm{ref}$) can be implemented in several ways, for example through direct adjustment, PI control, or incremental steps, with or without rate limiting.

\item \textit{Reactive power coordination:}  
The reactive power $Q_g$ is assumed to be regulated through conventional AC voltage or droop control, with $d$-axis priority. It is then constrained by the most restrictive capability limit (\ref{eq:Qg_max}).
\end{enumerate}

An advantage of the proposed method is that $P_g^*$ can be calculated analytically and updated in real time, without requiring global optimisation. The computational burden is therefore low and suitable for fast corrective control. Regardless of implementation, estimating the setpoint $P_g^*$ that maximises loadability as reactive margins approach depletion is essential for a SIPS targeting long-term voltage instability. Unlike EPC against frequency excursions, whether to increase or decrease $P_g^\mathrm{ref}$ is not as straightforward in the voltage-stability context, as both $P_g$ and $Q_g$ provide support, but their relative effectiveness depends on the system operating point.

\begin{figure}[htbp!]
    \centering
    \includegraphics[width=\linewidth, trim = 250 10 275 5, clip]{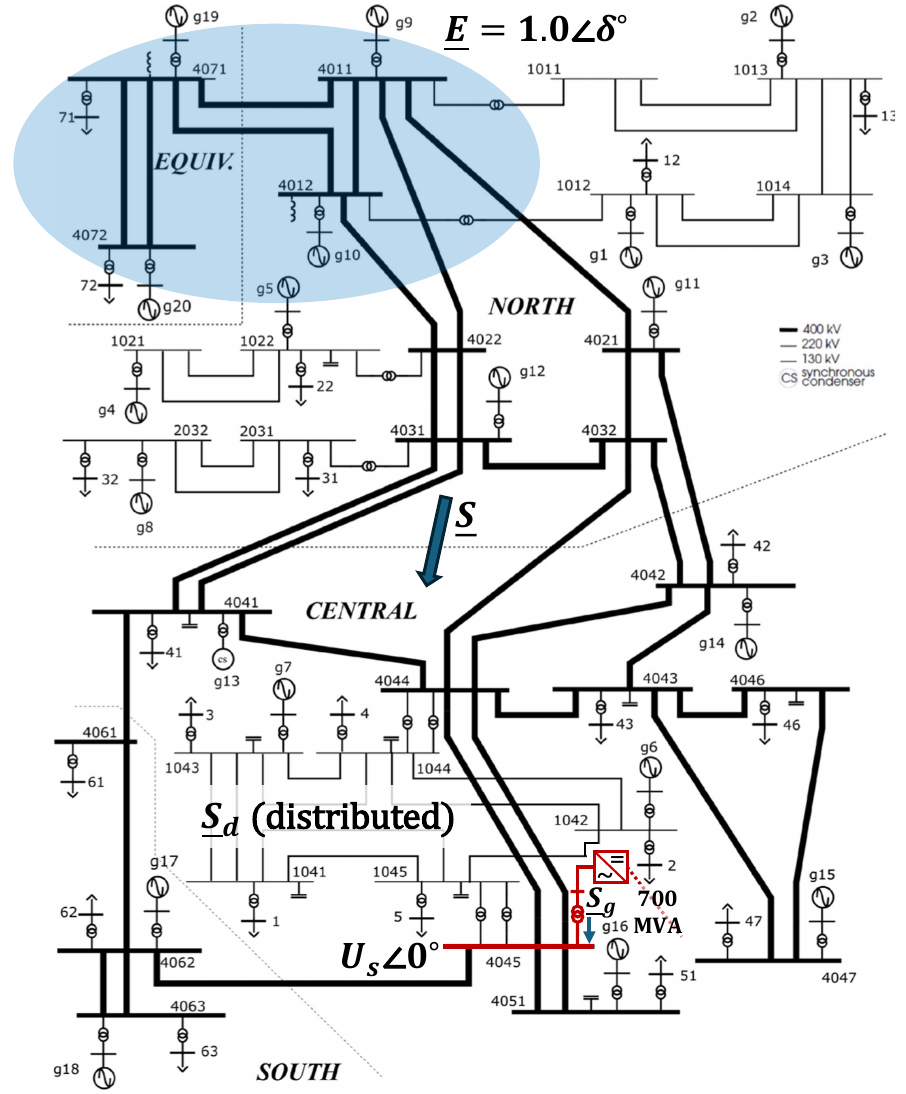}
    \caption{Single line diagram of the modified Nordic Test System with highlights, including the $700~\mathrm{MVA}$ VSC-HVDC link connected to bus 4045 ($\underline{U}_s = U_s\angle0^\circ$), and the remote bus cluster approximating $\underline{E}=1.0 \angle \delta^\circ$. $\underline{S}_d$ is the aggregate load in the Central area, $\underline{S}$ is the North $\rightarrow$ Central tie-line transfer reaching Central, and $\underline{S}_g$ is the HVDC power injection at bus 4045.}
    \label{fig:NordicSLD}
\end{figure}

In a real, meshed system, the remote node $\underline{E}$ in Fig.~\ref{fig:vsc_equivalent} does not exist explicitly. Nevertheless, $\delta$ can be obtained from a WAMS or state estimator by selecting physical bus proxies to represent $\underline{E}$, as done for the cluster in Fig.~\ref{fig:NordicSLD}. Preliminary results indicate that the REI (Radial, Equivalent, Independent) network reduction technique can be used to make this selection systematic, by having the REI generator bus represent $\underline{E}$ \cite{oatts1990application}. A detailed treatment of this is left for future work.

\section{Case Study and Validation}
\label{sec:case_study}
To validate the proposed active power setpoint expression in (\ref{eq:Pg_set}) and quantify the benefit, we perform dynamic phasor (RMS) simulations in the DIgSILENT PowerFactory implementation \cite{ospina2017implementation} of the Nordic Test System -- a standard IEEE benchmark system for voltage stability analysis \cite{vancutsem2020test}.

\subsection{Simulation Setup}

A VSC-HVDC model based on the NordBalt link was imported from the Northern European AC/DC Power System Model \cite{dijokas2021northern} and connected to the Nordic Test System. The HVDC model represents a $700~\mathrm{MVA}$, $\pm300~\mathrm{kV}$ bipole link connected to the $400~\mathrm{kV}$ Nordic HV grid through a $300/400~\mathrm{kV}$ step-up transformer (no tap-changer). The model includes a DC cable, converter and transformer to the asynchronous, continental European system (modelled as an infinite bus). The two converter stations operate in a master--slave configuration, with the VSC on the continental side regulating DC voltage while the one on the Nordic side directly controls $P_g$. The converter models are equipped with reference current ($i_d^\mathrm{ref},i_q^\mathrm{ref}$) limiters, which are set with $I_{c_{\max}} = 1.0~\mathrm{p.u.}$ and $d$-axis priority. The converter model, however, has no PWM-saturation limit. Therefore, an overvoltage limiter based on (\ref{eq:volt_lim}) is added to the outer control loop for $q$-axis current on the Nordic side, with $U_{c_{\max}} = 1.2~\mathrm{p.u}$. The converter (phase reactor) and transformer are set with the same reactances as in Table~\ref{tab:conv_data}, while resistances and filters are omitted.

The VSC-HVDC step-up transformer is connected to bus 4045, which represents $\underline{U}_s$, in the net-importing Central area. Reactive shunt compensation is added at this bus so that $U_s = 1.0~\mathrm{p.u.}$ for $(P_g, Q_g) = (700~\mathrm{MW},0~\mathrm{Mvar})$, i.e. nominal active power. To determine loadability for different setpoints $P_g^\text{ref}$, the VSC reactive power control mode is set to AC voltage control with $U_s^\mathrm{ref} = 1.0~\mathrm{p.u.}$ and a load increase in Central is simulated in the form of a slow load ramp, similar to the one in \cite{vancutsem2015test}. The loads are modelled using an exponential model ($P_d \propto V$, $Q_d \propto V^2$), and the load-scaling coefficients are increased such that, at nominal load voltages $V=V_0$, the aggregate active load in Central ramps at $0.15~\mathrm{MW/s}$. On-load tap-changers (OLTCs) are active on both the $130/20~\mathrm{kV}$ distribution transformers and the $400/130~\mathrm{kV}$ transmission transformers, maintaining load voltages (representing $V$ in Fig.~\ref{fig:vsc_equivalent}) close to nominal throughout the event. Frequency regulation is implemented only in the North and Equiv. areas.

\subsection{Results and Analysis}

\begin{figure}[htbp!]
    \centering
    \includegraphics[width=0.8\linewidth, trim = 0 5 0 5, clip]{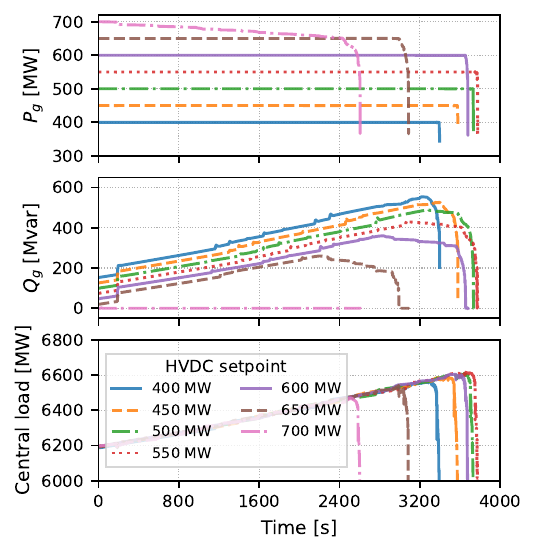}
    \caption{$P_g$, $Q_g$ and Central area total active load $P_d$ throughout the load ramp event for different HVDC setpoints $P_g^\text{ref}$. The setpoint that provides the highest loadability and thus extends the time to collapse the most is $P_g^\text{ref} = 550~\mathrm{MW} \approx 0.8~\mathrm{p.u.}$, closely followed by $500$ and $600~\mathrm{MW}$.}
    \label{fig:PQ_and_P_central}
\end{figure}

\begin{figure}[htbp!]
    \centering
    \includegraphics[width=0.9\linewidth, trim = 5 5 5 5, clip]{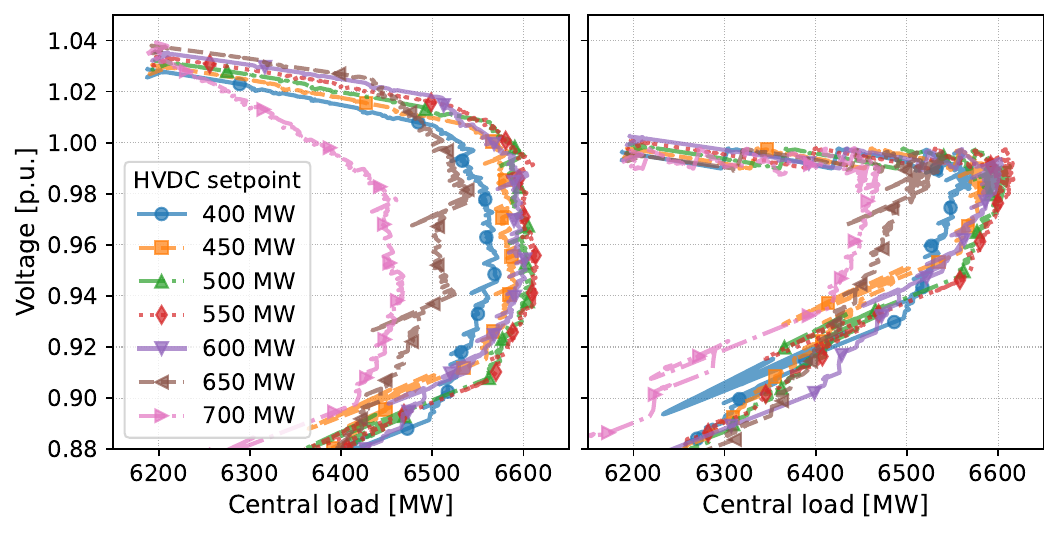}
    \caption{PV curves of bus voltage versus total active power load in Central for different HVDC setpoints $P_g^\text{ref}$. Left: bus 4044 ($400~\mathrm{kV}$), right: bus 1044 ($130~\mathrm{kV}$). The setpoint yielding the highest loadability is $P_g^\mathrm{ref} = 550~\mathrm{MW}$.}
    \label{fig:PV}
\end{figure}

\begin{figure}[htbp!]
    \centering
    \includegraphics[width=0.8\linewidth, trim = 0 5 0 5, clip]{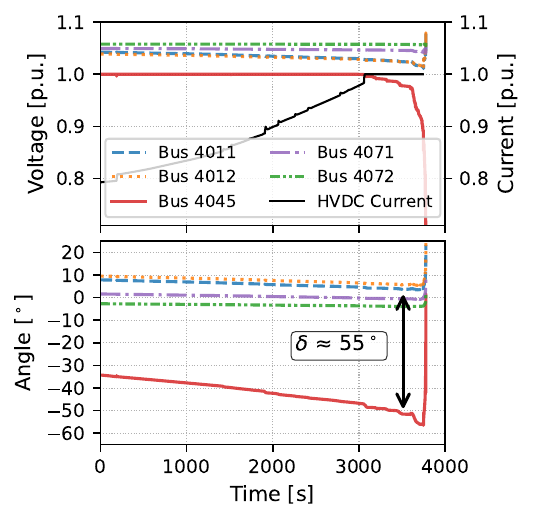}
    \caption{Evolution of voltage magnitudes and angles with $P_g^\mathrm{ref} = 550~\mathrm{MW}$. When the current limit binds and voltage collapse approaches, $\delta$ is $50$--$60^\circ$.}
    \label{fig:delta_550}
\end{figure}

Simulations were run for seven cases with different HVDC setpoints $P_g^\mathrm{ref} \in \{400, 450,\ldots,700\}~\mathrm{MW}$, and selected results are shown in Figs.~\ref{fig:PQ_and_P_central}--\ref{fig:delta_550}. In Fig.~\ref{fig:PQ_and_P_central}, HVDC power output and Central load are shown over time for all seven cases. With $P_g^\mathrm{ref} = 700~\mathrm{MW}$, the HVDC output is current-limited from the start, and as the load ramp progresses, no reactive power can be provided, causing the earliest voltage collapse. Reducing the active power setpoint $P_g^\text{ref}$ by $100~\mathrm{MW}$ ($700~\mathrm{MW} \rightarrow 600~\mathrm{MW}$), the loadability $P_d$ can be increased by about $150~\mathrm{MW}$. Both $P_d$ and the time to collapse then reach a maximum before decreasing slowly again for $P_g^\text{ref}$ below $500~\mathrm{MW}$. This corroborates that the reasoning and results from the smaller system (see $U_s = 1.0~\mathrm{p.u.}$ in Fig.~\ref{fig:Pd_Pg}) can be transferred to a larger, more complex system.

PV curves from the same simulations are shown in Fig.~\ref{fig:PV} for the two Central buses 4044 ($400~\mathrm{kV}$) and 1044 ($130~\mathrm{kV}$). The curves show that the load ramp causes a declining voltage profile at the $400~\mathrm{kV}$ transmission level (with the exception of $U_s$), while the OLTCs maintain voltages at the $130~\mathrm{kV}$ subtransmission level close to nominal until collapse. The highest loadability $P_d$ is found as the nose point of each curve.

The first of the two abrupt voltage drops observed for the case $P_g^\mathrm{ref} = 650~\mathrm{MW}$ in Fig.~\ref{fig:PV} can be attributed to the HVDC current limit: as the load ramp depresses $U_{s}$ below $650/700 \approx 0.93~\mathrm{p.u.}$, $Q_g$ rapidly drops to zero. This is consistent with the circle's vertical tangent ($dQ_g/dP_g \rightarrow -\infty$ as $P_g-P_0^{(i)} \rightarrow r_i$). The second drop occurs upon OEL activation in generator g14.

The voltage angle difference $\delta$ between the strong grid $\underline{E}$ and the HVDC connection bus $\underline{U}_s$ is measured from a cluster of strong buses in the North and Equiv. areas (see Fig.~\ref{fig:NordicSLD}). Near voltage collapse, when a SIPS would likely be triggered, $\delta$ is in the interval $50$--$60^\circ$ (see Fig.~\ref{fig:delta_550}). Although this estimate is not exact, it is larger than $\delta_i$, which for the present parameters evaluates to $29$--$49^\circ$ for $U_s = 0.95$--$1.0~\mathrm{p.u.}$ (\ref{eq:deltai_deltav}). Consequently, we expect to be in the current-limited regime ($\delta > \delta_i$), with the margin increasing as $U_s$ decreases. With $\delta = 50$--$60^\circ$ and $U_s = 0.95$--$1.0~\mathrm{p.u.}$, the optimal $P_g^*$ can be calculated through (\ref{eq:Pg_set}) to be in the interval $510\textrm{--}610~\mathrm{MW}~(0.73\textrm{--}0.87~\mathrm{p.u.})$. These are precisely the setpoints $P_g^\mathrm{ref}$ that are observed to provide the highest loadability in Figs.~\ref{fig:PQ_and_P_central}--\ref{fig:PV}, further validating the method.

Adjusting the shunt compensation so that $U_s = 1.05~\mathrm{p.u.}$ for $(P_g, Q_g) = (700~\mathrm{MW},0~\mathrm{Mvar})$ and setting $U_s^\mathrm{ref} = 1.05~\mathrm{p.u.}$, the setpoint yielding the highest loadability increases accordingly. As expected, the voltage limit is also more frequently the first to bind. However, these results should be interpreted with caution: the simulations use fixed setpoints $P_g^\mathrm{ref}$, whereas the optimum $P_g^*$ changes with $U_s$ and $\delta$. Full validation requires $P_g^*$ to be implemented and updated at each control interval through an EPC framework, as described in Section~\ref{sec:epc_scheme}.

\section{Limitations and Practical Considerations}

The proposed method relies on simplifying assumptions and measured quantities that may be subject to uncertainty, particularly the voltage magnitude $U_s$ and angle difference $\delta$. Therefore, this section examines the sensitivity of the optimal setpoint $P_g^*$ and loadability $P_d$ to these variables, and discusses practical aspects related to implementation and operation.

\subsection{Sensitivity Analysis}

\begin{figure}[htbp!]
    \centering

    \subfloat[$U_s = 0.9~\mathrm{p.u.}$]{%
        \includegraphics[width=0.32\linewidth, trim = 0 5 0 5, clip]{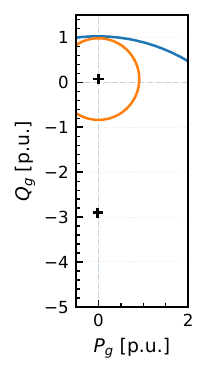}%
    }
    \hfill
    \subfloat[$U_s = 1.0~\mathrm{p.u.}$]{%
        \includegraphics[width=0.32\linewidth, trim = 0 5 0 5, clip]{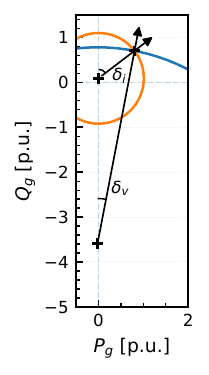}%
    }
    \hfill
    \subfloat[$U_s = 1.1~\mathrm{p.u.}$]{%
        \includegraphics[width=0.32\linewidth, trim = 0 5 0 5, clip]{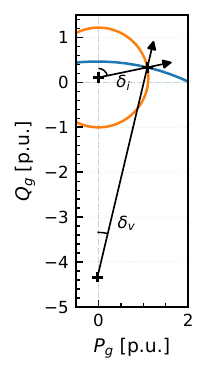}%
    }

    \caption{VSC-HVDC $PQ$-capability chart (current and voltage limits) at the system bus for different voltage magnitudes $U_s$.}
    \label{fig:PQ_different_Us}
\end{figure}

As $U_s$ changes, both the position and size of the current and voltage limits shift (see Fig.~\ref{fig:PQ_different_Us}), also affecting the transition angles $\delta_i$ and $\delta_v$. The optimal setpoint $P_g^*$ is thus sensitive to voltage variations, particularly in the intersection regime ($\delta_v < \delta < \delta_i$). However, accurate local voltage measurements are essential for almost all HVDC control strategies. It is therefore reasonable to assume that $U_s$ is available with sufficient accuracy, possibly from the local PMU.

The angle difference $\delta$ is inherently less certain than $U_s$, since the remote node $\underline{E}$ in Fig.~\ref{fig:vsc_equivalent} does not exist explicitly in real, meshed systems. Instead, $\underline{E}$ must be approximated, for instance by physical bus proxies. This analysis thus examines deviations in $\delta$ on the order of $\pm10^\circ$ and shows that the effect on the computed setpoint and loadability remains acceptable.

Examining (\ref{eq:Pg_set}) and Fig.~\ref{fig:ss_and_optimalPQ}, we note that $P_g^*$ is independent of $\delta$ in the intersection regime ($\delta_v < \delta < \delta_i$). In the current-limited regime, we typically have $\delta > 45^\circ \implies dP_g^*/d\delta = r_i\cos{\delta}<1$, partially mitigating propagation of the error. The optimal setpoint is unlikely to lie in the voltage-limited regime $(\delta < \delta_v)$ in practice: $\delta$ is rarely below $\delta_v$ near collapse, and the reactive power gained by reducing $P_g$ below $P_x$ is small.

\begin{figure}[htbp!]
    \centering
    \includegraphics[width=0.85\linewidth, trim = 0 5 0 5, clip]{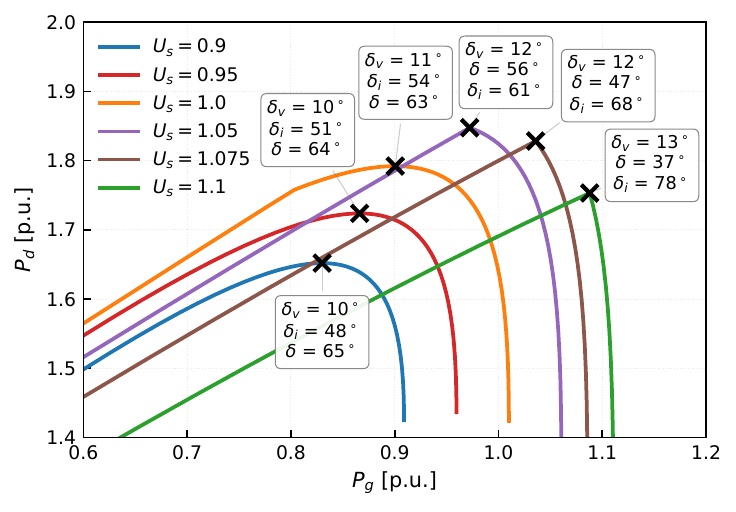}
    \caption{$P_d$ as a function of $P_g$ for different voltage magnitudes $U_s = V/n$ through (\ref{eq:Pd})--(\ref{eq:Q_bal}). $Q_g$ is parametrised as the most restrictive limit according to (\ref{eq:Qg_max}). $P_g^*$ is annotated at the peak of each curve, with the corresponding $\delta$ and transition angles $\delta_i, \delta_v$. $E = 1.0~\mathrm{p.u.}$, $V = 1.0~\mathrm{p.u.}$ and $X = 1.0~\mathrm{p.u.}$}
    \label{fig:Pd_Pg}
\end{figure}

While the above discussion focuses on the sensitivity of $P_g^*$, it is equally important to assess how deviations in $P_g$ affect the loadability $P_d$, since this determines how precisely $P_g^*$ must be tracked. This is done by expressing $P_d$ as a function of $P_g$ (see (\ref{eq:Pd})--(\ref{eq:Q_bal}) in the appendix), with $Q_g$ parametrised along the most restrictive limit (\ref{eq:Qg_max}). The resulting relationship is illustrated in Fig.~\ref{fig:Pd_Pg} for different voltage levels $U_s$.

When $P_g - P_0^{(i)} \rightarrow r_i$, we note that $dP_d/dP_g \rightarrow -\infty$ and the loadability $P_d$ can be significantly increased by only a slight reduction in $P_g$. Furthermore, when $P_g^*$ is in the current-limited regime ($\delta > \delta_i$), which in the figure is the case for $U_s \leq 1.0~\mathrm{p.u.}$, the optimum is on a relatively flat plateau, giving some leeway for errors in $P_g^*$ and thus also measurement errors in $U_s$ and $\delta$. In contrast, when $P_g^*$ is in the intersection regime ($\delta_i > \delta > \delta_v$), which is the case for $U_s > 1.05~\mathrm{p.u.}$, the optimum is at an edge, sharper the higher $U_s$ becomes. Fortunately, $P_g^*$ is then independent of $\delta$. 

Based on the curve gradients in Fig.~\ref{fig:Pd_Pg}, it is more detrimental to overestimate $P_g^*$ than to underestimate it. To maintain a non-zero reactive power margin, it is thus recommended to keep a small margin $P_g^\mathrm{ref} - P_0^{(i)} <~r_i$. It is also clear that $P_d$ can be increased by voltage setpoint adaptation of $U_s$, with a maximum for $U_s = 1.05~\mathrm{p.u.}$ in Fig.~\ref{fig:Pd_Pg}. Finally, it is worth noting that $P_d$ is dependent on $E$, $X$ and $Q_d$, while $P_g^*$ is not.

\subsection{Practical Considerations}

A key difference between the VSC-HVDC voltage limit (\ref{eq:volt_lim}) and the analogous OEL in a synchronous generator is that the OEL shifts only marginally with changes in terminal voltage, due to the relatively large field voltage (typically $E_{fd}>2.0~\mathrm{p.u.}$) and synchronous reactance (typically $X_d>1.0~\mathrm{p.u.}$) \cite{vancutsem2015test}. In contrast, the VSC-HVDC voltage limit (\ref{eq:volt_lim}) is much more sensitive to variations in $U_s$, due to the smaller maximum converter voltage $U_{c_{\max}}$ ($\approx 1.2~\mathrm{p.u.}$) and phase reactance $X_c$ ($\approx 0.1~\mathrm{p.u.}$). This can also be seen by comparing Fig.~\ref{fig:PQ_different_Us} with Figs.~2.12--2.13 (pp.~31--32) in \cite{vancutsem2015test}.

To avoid excessive reduction of the reactive power capability by the VSC-HVDC voltage limit, it is important to maintain a positive voltage difference $U_{c_{\max}} - U_s$, ideally larger than $(X_c + X_{tf})I_{c_{\max}}$ so that the voltage limit is less restrictive than the current limit for all $P_g$ (in the lossless, filterless case). This can be achieved through reactive power compensation at the system ($\underline{U}_s$) or filter bus ($\underline{U}_f$), by equipping the step-up transformer ($\underline{Z}_{tf}$) with OLTCs, or by increasing $U_{c_{\max}}$ in the design stage through, e.g., third-harmonic injection PWM or a higher $U_{DC}/U_{AC}$-ratio, thereby avoiding PWM saturation. 

This work focuses on slow outer-loop dynamics and EPC behaviour when converter limits are binding, effectively rendering the VSC-HVDC station as a $PQ$-bus. The distinction between grid-forming and grid-following control is therefore expected to be of limited relevance in this context.

\section{Summary and Conclusions}

This paper has presented an analytical method for determining the optimal active power setpoint of a VSC-HVDC station under combined current and voltage constraints, with the objective of maximising loadability under voltage-stressed conditions. By exploiting the geometric structure of converter capability limits in the $P$--$Q$ plane, a closed-form expression for the optimal setpoint $P_g^*$  was derived as a function of the wide-area voltage angle difference $\delta$ and local bus voltage $U_s$.

The derived expression can be integrated into a SIPS targeting long-term voltage instability. A key advantage is that $P_g^*$ requires only two input signals and can be updated in real time without global optimisation, reducing computational burden and increasing robustness. The expression was corroborated through dynamic simulations on the Nordic Test System, where the active power setpoint $P_g^\mathrm{ref}$ providing the highest loadability was consistent with the analytically predicted $P_g^*$. A sensitivity analysis further demonstrated that uncertainties in the wide-area angle difference $\delta$ propagate only weakly to the loadability $P_d$ in the regimes where $\delta$ is required, supporting practical feasibility.

Closed-loop implementation and dynamic evaluation of the proposed method within an EPC framework are left for an extended version. Likewise, coordination between the proposed method and other remedial actions targeting voltage instability, such as under-voltage load shedding, tap-changer blocking, and voltage setpoint adjustment, warrants further investigation.

\appendix
\subsection{Derivation of Eq. (\ref{eq:opt_any_centre})}
Operating at either of the circular limits in (\ref{eq:curr_lim}) or (\ref{eq:volt_lim}) yields:
\begin{equation}
\label{eq:circ_lim}
(P_g - P_0)^2 +  (Q_g - Q_0)^2=r^2.
\end{equation}

Active power balance at the system bus is given by: 
\begin{equation}
\label{eq:Pd}
P_d = P_g+\dfrac{EV}{nX}\underbrace{\sqrt{1-\cos^2{\delta}}}_{=\sin\delta}.
\end{equation}

Reactive power balance at the system bus, in turn, gives:
\begin{equation}
\label{eq:Q_bal}
\begin{aligned}
&\dfrac{EV}{nX}\cos\delta-\dfrac{V^2}{n^2X}+Q_g = 0 \overset{\scriptstyle (\ref{eq:circ_lim})}{\implies} \\
&\cos\delta = \dfrac{V}{En} - \dfrac{nX}{EV} 
\underbrace{\left(Q_0+\sqrt{r^2-(P_g-P_0)^2}\right)}_{=Q_g}.
\end{aligned}
\end{equation}

Differentiating (\ref{eq:Pd}) w.r.t. $P_g$ and substituting $\cos \delta$ from (\ref{eq:Q_bal}) now yields:
\begin{equation}
\dfrac{\partial P_d}{\partial P_g} = 1 - \dfrac{(P_g-P_0)\cos\delta}{(Q_g-Q_0)\sin\delta}.
\end{equation}

Setting $\partial P_d / \partial P_g = 0$, we have
\begin{equation}
\dfrac{Q_g-Q_0}{P_g-P_0} = \dfrac{1}{\tan\delta} \notag. \quad \square
\end{equation}

The second derivative is negative, confirming that this stationary point corresponds to a maximum. As in \cite{johansson1997avoiding}, it is assumed that $E$ and $X$ are constant and that $n$ changes slowly. The converter parameters, including $I_{c_{\max}}$ and $U_{c_{\max}}$, are also assumed constant.

\bibliography{references.bib}
\bibliographystyle{ieeetr}

\end{document}